\documentclass[twocolumn]{aastex62}

\usepackage{graphicx}
\usepackage{soul,xcolor}
\usepackage{amsmath}
\usepackage{mathtools}
\usepackage{multirow}
\usepackage{hyperref}
\usepackage{soul}

\graphicspath{ {figures/} }  

\usepackage{natbib}
\usepackage{CJKutf8}
\newcommand{\cntext}[1]{\begin{CJK*}{UTF8}{bsmi}#1\end{CJK*}}

\received{December 3, 2023}
\revised{\today}
\submitjournal{ApJ}

\shortauthors{Wang, Chen, \& Pan}

\begin{document}

\title{Type Ia Supernova Progenitors and Surviving Companions within the Symbiotic Channel}

\newcommand*{\NTHUP}{Department of Physics, National Tsing Hua University, Hsinchu 30013, Taiwan}
\newcommand*{\NTHUA}{Institute of Astronomy, National Tsing Hua University, Hsinchu 30013, Taiwan}
\newcommand*{\CICA}{Center for Informatics and Computation in Astronomy, National Tsing Hua University, Hsinchu 30013, Taiwan}
\newcommand*{\CTC}{Center for Theory and Computation, National Tsing Hua University, Hsinchu 30013, Taiwan}
\newcommand*{\NCTS}{Physics Division, National Center for Theoretical Sciences, Taipei 10617, Taiwan}
\newcommand*{\TSMC}{Taiwan Semiconductor Manufacturing Company, Hsinchu, Taiwan}
\newcommand*{\UT}{Department of Astronomy, The University of Texas at Austin, 2515 Speedway, Stop C1400, Austin, Texas 78712-1205, USA}

\author{Yu-Hui Wang (\cntext{王榆惠})}
\altaffiliation{These authors contributed equally to this work.}
\affiliation{\NTHUP} \affiliation{\TSMC}

\author[0000-0003-1640-9460]{Hsin-Pei Chen (\cntext{陳昕霈})}
\altaffiliation{These authors contributed equally to this work.}
\affiliation{\NTHUP} \affiliation{\NTHUA} \affiliation{\UT}

\author[0000-0002-1473-9880]{Kuo-Chuan Pan (\cntext{潘國全})}
 \affiliation{\NTHUP} \affiliation{\NTHUA} 

\begin{abstract}
The symbiotic channel of Type Ia supernovae progenitors is crucial for explaining the observed circumstellar material in some Type Ia supernovae.
While extensive numerical and observational efforts have been dedicated to exploring the progenitor system, limited emphasis has been placed on studying the surviving companions arising from the symbiotic channel.
In this paper, we present a numerical study of the symbiotic systems using {\tt MESA} as potential Type Ia supernova progenitors. 
We conduct 1260 binary stellar evolution simulations, over a wide range of parameters, incorporating the optically thick wind model developed by Hachisu et al.,
and predict the post-impact evolution of these surviving companions.
We classify four types of progenitor systems based on the evolutionary stage of the companion at the onset of the explosion: red giant companions, with or without prior helium flash events, and asymptotic giant branch companions, with or without the thermal pulsing phase.
After the SN impact, a blue dwarf star with either a helium or carbon-oxygen core is left behind. However, if a small portion of the envelope ($\gtrsim$ 0.3\%) remains on the core of the surviving companion, the overall post-supernova evolution may remain similar to its pre-explosion state, albeit slightly fainter, making observation a challenging endeavor.
\end{abstract}


\section{INTRODUCTION}

Type Ia supernovae (SN~Ia) are thermonuclear explosions of accreting carbon-oxygen (CO) white dwarfs (WD).
Because of their homogeneous formation mechanism, SNe~Ia are pivotal cosmological tools as the standardizable candles \citep{1992ARAA..30..359B, 1998AA...331..815T, 2011ApJ...731..120M}.
However, the mystery regarding their progenitor systems, crucial for uncovering their intrinsic variations, remains unsolved \cite[e.g.][]{1995PASP..107.1019B, 2000ARAA..38..191H, 2012NewAR..56..122W, 2019NewAR..8701535S, 2020IAUS..357....1R, 2023RAA....23h2001L}. 
Two widely investigated SNe~Ia progenitor scenarios are the single-degenerate scenario \cite[SD scenario,][]{1973ApJ...186.1007W, 1982ApJ...257..780N, 1984ApJS...54..335I, 1985ASSL..113....1P} and the double-degenerate scenario \cite[DD scenario,][]{1984ApJS...54..335I, 1984ApJ...277..355W}. 
The SD scenario considers mass transfer during the binary interaction of a WD and a non-degenerate companion, which could be a main-sequence (MS) star, a helium (He) star, a red giant (RG) star, or an asymptotic giant branch (AGB) star. 
The DD scenario suggests that SNe~Ia result from the mass transfer and merger of two WDs.

One direct method to distinguish between these two scenarios is to search for the non-degenerate companions in SN remnants (SNR). The DD scenario does not expect a surviving companion after the explosion, whereas several numerical studies have suggested that the surviving companion in the SD scenario could be observable \citep{2010ApJ...715...78P, 2012ApJ...750..151P, 2012ApJ...760...21P, 2013ApJ...773...49P, 2014ApJ...792...71P, 2013ApJ...778..121L, 2022ApJ...928..146L, 2021MNRAS.500..301L, 2019ApJ...887...68B, 2020ApJ...898...12Z, 2022ApJ...933...65Z, 2021AA...646L...8N, 2022ApJ...933...38R}. 
However, despite a few candidates \cite[e.g.][]{2014ApJ...792...29F, 2015ApJ...799..101P, 2023ApJ...947...90R}, no surviving companion of SN~Ia systems has been conclusively confirmed \cite[e.g.][]{2013ApJ...774...99K, 2018ApJ...862..124R, 2018MNRAS.479.5696K, 2019ApJ...886...99L, 2023ApJ...950L..10S}. 
Type Iax SN~2012Z is the only peculiar SN~Ia with a pre-SN detection of the progenitor system
\citep{2014Natur.512...54M, 2022ApJ...925..138M}, indicating a WD+He star origin, while the surviving companion has not been observed due to the SN's enduring high brightness.
There even emerged a potential candidate, SN~2020hvf, as the result of a double-detonation explosion, with an inflated companion spectroscopically observed \citep{2023arXiv230611788S}.

Another method to discriminate between progenitor scenarios is observing the signatures arising from the interaction between the SN ejecta and the circumstellar material (CSM). 
CSM is expected to form during the late-time pre-SN mass transfer in the SD scenario. By contrast, the mass transfer process ceases a long time ago before the merger-induced SN explosion in the DD scenario. 
The time-varying Na ID absorption line, suggesting ejecta-CSM interaction,
have been observed in tens of SNe~Ia \cite[SNe~Ia-CSM,][]{2007Sci...317..924P, 2013ApJS..207....3S, 2023ApJ...948...52S}.

In the context of the SD scenario, the symbiotic channel, which involves a WD accreting matter from a RG companion star, stands out as a promising pathway for generating dense CSM before the SN events \cite[e.g.][]{1999ApJ...522..487H, 2008ApJ...679.1390H, 2009MNRAS.396.1086L, 2016MNRAS.457..822B, 2019AA...622A..35L}.
Theoretically, the optically thick wind model \citep{1996ApJ...470L..97H, 1999ApJ...522..487H} is a classic of the symbiotic channel because it overcame a primary challenge in SD scenario theories: how to stably grow the WD star until it reaches the Chandrasekhar mass $M_{\rm Ch}$.
In addition to the typical Roche-lobe overflow (RLOF) mechanism, \cite{1996ApJ...470L..97H} ensured the steady mass growth by introducing an optically thick wind blowing from the WD during mass accretion, and \cite{1999ApJ...522..487H} expanded the possible parameter space leading to SNe~Ia by considering a mass-stripping effect caused by the WD wind interacting with the envelope of the donor.

Later on, issues related to the assumptions of symmetric accretion and stellar wind have been discussed.
For example, \cite{2009MNRAS.396.1086L} proposed a model that takes into account aspherical RG wind and identified a broader range of possible parameter space than the previous work. 
\cite{2021MNRAS.503.4061W} also discussed the effect of the aspherical RG wind on the geometric structure of the CSM but did not find a strong correlation between the wind opening angle and the final CSM structure.
Alternatively, without considering the mass-stripping effect of the WD, \cite{2011ApJ...735L..31C} introduced the concept of accretion from the tidally enhanced stellar wind of the RG and successfully expanded the potential parameter space of symbiotic channel as well.
\cite{2019MNRAS.485.5468I} also considered RG-wind accretion in their population synthesis simulations of symbiotic system V407~Cyg.
Their argument posited that V407~Cyg-like systems do not evolve into SNe~Ia when adopting the standard evolution model. However, they demonstrated that considering RG-wind accretion allows these systems to potentially become progenitors for SNe~Ia.

The symbiotic channel has been examined by the observational signatures of a few SNe~Ia-CSM and some known symbiotic binaries \citep{2023ApJ...948...52S}.
For example, SN~2006X \citep{2007Sci...317..924P, 2008AstL...34..389C} and PTF11kx \citep{2012Sci...337..942D, 2013MNRAS.431.1541S} were detected to have dense CSM; however, their origins from symbiotic systems remain a topic of debate.
SN~2020eyj \citep{2023Natur.617..477K} is the first SN~Ia-CSM to have radio detection, indicating the presence of dense CSM, and the existence of narrow He emission lines suggests a WD+He star origin.
SN~1604, also known as the Kepler supernova, was proposed by \cite{2012AA...537A.139C} to have a symbiotic progenitor, based on the morphological evolution of the SNR in their hydrodynamics simulations. However, no sub-giant or RG surviving companion has been found \citep{2014ApJ...782...27K, 2018ApJ...862..124R}.
RS~Ophiuchi, the most widely investigated symbiotic recurrent nova system, exhibits recurrent nova outbursts with a period of $\sim$ 20 years and contains a growing WD close to $M_{\rm Ch}$ \citep{2007ApJ...659L.153H, 2017ApJ...847...99M}.
Because the behavior of the absorption lines of RS~Ophiuchi shares similarity with that of SN 2006X, it is believed to be an upcoming SN~Ia candidate \citep{2011AA...530A..63P, 2016MNRAS.457..822B}.
Moreover, symbiotic recurrent novae like T~CrB \citep{2019ApJ...886...99L} and V3890~Sgr \citep{2021MNRAS.504.2122M} have observational properties similar to RS~Ophiuchi, making them candidates as progenitors for SNe~Ia as well.

Beyond the symbiotic systems, an interacting binary with an AGB companion, positioned at a latter evolutionary stage than a RG companion, stands as another promising candidate for producing CSM through the strong stellar wind emitted by the AGB star.
For instance,
SN~2002ic is the very first SN~Ia-CSM classified by \cite{2003Natur.424..651H}, which proposed a WD$+$AGB origin based on the strong H absorption line from the ejecta-CSM interaction, indicating significant mass loss before the SN. Note that there has been ongoing debate about classifying a SN with strong H detection as a SN~Ia event \cite[e.g.][]{2006ApJ...653L.129B}.
Another candidate is the Type Iax SN~2008ha, in which \cite{2014ApJ...792...29F} detected an AGB-like source four years after the SN event, while this source can also be explained by an extremely inflated CSM/SNR of a single WD.
Recently, \cite{2023RAA....23g5010L} conducted numerical simulations covering a wide parameter space of WD$+$AGB progenitor systems, considering RLOF and AGB-wind accretion. They provided a parameter space for forming SNe~Ia within this channel and suggested the nova-AGB system AT~2019qyl \citep{2021ApJ...920..127J} as a potential candidate.

Regarding the birthrate of SNe~Ia through the symbiotic channel, \cite{2009MNRAS.396.1086L} estimated a Galactic birthrate ranging between $1.03 \times 10^{-3}$ and $2.27 \times 10^{-5}$ $\rm yr^{-1}$ derived from their binary evolution simulations considering an aspherical RG wind model. 
\cite{2023BASBr..34..170L} proposed a population of the symbiotic channel in the Local Group galaxies ranging from $1.69 \times 10^3$ to $3.23 \times 10^4$, using both empirical and theoretical approaches. 
They also suggest that the symbiotic channel constitutes 0.5-8\% of the Galactic SNe~Ia.

In this study, we conduct a series of one-dimensional stellar evolution simulations using an ad-hoc scheme that includes the physics of binary interactions and supernova impact.
We re-implement the optically thick wind model and the initial parameter space of the SN~Ia progenitor systems as described by \cite{1996ApJ...470L..97H, 1999ApJ...522..487H} into the open-source stellar evolution code {\tt MESA}\footnote{http://mesa.sourceforge.net/} (Modules for Experiments in Stellar Astrophysics; \citealt{2011ApJS..192....3P, 2013ApJS..208....4P, 2015ApJS..220...15P, 2018ApJS..234...34P, 2019ApJS..243...10P}; version: r12115; mesasdk version: 200301). 
While \cite{1996ApJ...470L..97H, 1999ApJ...522..487H} primarily focus on the final types of SN~Ia progenitors, their work lacks a detailed discussion regarding the properties of companion stars. Furthermore, \cite{1999ApJ...522..487H} did not actually conduct binary stellar evolution simulations, but rather used the empirical formulae from \cite{1983ApJ...270..678W}.
Therefore, our approach using {\tt MESA} should provide more self-consistent simulations of the evolution of companion stars.
In our study, our focus shifts to companion stars, encompassing their final evolutionary stages just before the SN event and their post-impact evolution as surviving companions.
In addition, we find that the companion stars could evolve into He-rich RG stage that has experienced a He-core flash (hereafter referred to as the post-He-flash RG stage) or AGB stage at the onset of SN~Ia explosion within our considered parameter space, as \cite{1996ApJ...470L..97H, 1999ApJ...522..487H} simply consider them to be in the RG stage.

The paper is structured as follows: Section~\ref{sec:method} describes the numerical methods and setups for binary evolution and post-SN companion evolution. Section~\ref{sec:res} presents the binary evolution results, compares them with \cite{1999ApJ...522..487H}, and examines the post-impact evolution of the surviving companions under two scenarios: (1) a remaining stellar core and (2) a stellar core with some remaining envelope. We then explore the observability of these surviving companions in the symbiotic channel. 
Section~\ref{sec:dis} discusses the absence of detectable surviving companions in Type Ia supernova remnants, particularly in the symbiotic channel and post-He-flash RG models.
Section~\ref{sec:sum&con} summarizes our findings and conclusions.

\section{NUMERICAL METHODS}
\label{sec:method}

In this section, we briefly review the adopted theoretical model and describe the simulation setups.
The whole simulations contain two steps: Firstly, binary evolution simulations are generated by the 1D stellar evolution code {\tt MESA}. We adopt the optically thick wind model and mass-stripping effect from \cite{1996ApJ...470L..97H, 1999ApJ...522..487H}. 
Secondly, 
we develop a toy model in {\tt MESA} that effectively considers the envelope-stripping and SN-heating during the SN impact on the binary companions. Using this toy model, we perform the long-term evolution of the surviving companions in the SNR within the symbiotic channel. 

\subsection{Binary Evolution}
\label{sec:method_BE}

We generate binary stellar evolution models in {\tt MESA}, each consisting of a WD (described by a point mass) and a donor (evolved starting from a zero-age MS star), implementing the optically thick wind model with the mass-stripping effect \citep{1996ApJ...470L..97H, 1999ApJ...522..487H}. 
Here, we briefly summarize this model for the consistency of this paper.

The primary mechanism of the optically thick wind model is that when the mass ratio $q$ of the primary (the WD; its mass denoted as $M_{\rm WD}$) and the secondary (the companion star/donor; its mass denoted as $M_{\rm d}$) reaches $q = M_{\rm d}/M_{\rm WD} > 0.79$, or when the mass transfer rate $\dot{M}_{\rm t}$ exceeds the critical accretion rate, which is
\begin{equation}
    \dot{M}_{\rm cr} \approx 0.75 \times 10^{-6} \left(\frac{M_{\rm WD}}{\rm M_\odot} - 0.4 \right)~\rm M_\odot~{yr}^{-1},
\end{equation}
an optically thick wind starts blowing from the WD surface, stabilizing the mass transfer and maintaining the accretion rate at $\dot{M}_{\rm cr}$.
The mass transfer rate is given by Equations (5)-(7) in \cite{1999ApJ...522..487H}.

During the WD wind phase, the mass accumulation ratio (denoted by $\eta$) of the H shell burning on the WD surface is given by the conditions of the WD accretion rate $\dot{M}_{\rm WD}$,
\begin{equation}
    \eta_{\rm H}=
 \begin{cases} 
          0, & \text{\rm for}~|\dot{M}_{\rm WD}| \leq \dot{M}_{\rm low}, \\
          1, & \text{\rm for}~\dot{M}_{\rm low} < |\dot{M}_{\rm WD}| < \dot{M}_{\rm cr}, \\
          (1-\dot{M}_{\rm wind}/\dot{M}_{\rm WD}), & \text{\rm for}~\dot{M}_{\rm cr} \leq |\dot{M}_{\rm WD}| \lesssim \dot{M}_{\rm high}, 
 \end{cases}
\end{equation}
where $\dot{M}_{\rm low} =\frac{1}{8} \dot{M}_{\rm cr} = 10^{-7} \rm ~M_{\odot}~yr^{-1}$ and $\dot{M}_{\rm high}=10^{-4}\rm ~M_{\odot}~yr^{-1}$ are the lowest and the highest mass accretion rates that maintain the WD wind, and $\dot{M}_{\rm wind}$ is the wind mass loss rate.
As the H shell burns and the He shell accumulates, He shell flashes occur under certain conditions, which are described by the He mass accumulation ratio, 
\begin{equation}
    \eta_{\rm He}=
 \begin{dcases} 
      1.05-0.175(\log \dot{M}_{\rm He}+5.35)^2, 
      & \\ 
      \qquad\qquad\text{\rm for} -7.3<\log \dot{M}_{\rm He} < -5.9, \\
      1\mathmakebox[3.3em][l], \text{\rm for}-5.9 \leq \log \dot{M}_{\rm He} < -5.0,
   \end{dcases} \label{eq_etaHe}
\end{equation}
where $\dot{M}_{\rm He}=\eta_{\rm H}|\dot{M}_{\rm t}|$ is the net helium mass accretion rate of the WD.

Meanwhile, the optically thick wind partially strips the donor's envelope and lowers the mass transfer rate, called the mass-stripping effect. 
The mass stripping rate is given by
\begin{equation}
    \dot{M}_{\rm str} = c_1 \, \dot{M}_{\rm wind},
\end{equation}
where the coefficient $c_1$ can be approximated by
\begin{equation}
    c_1 \approx \eta_{\rm eff} \left[ \frac{g(q)}{0.025}\right] \left( \frac{v}{1000 \, \rm km \, s^{-1}}\right)^2 \left( \frac{R_{\rm d}}{30 \, \rm R_{\odot}}\right) \left( \frac{M_{\rm d}}{2\, \rm M_{\odot}}\right)^{-1},
\end{equation}
$\eta_{\rm eff}=1$ is the efficiency of mass stripping, 
$g(q)$ is a geometrical factor of the donor's surface,
$v$ is the wind velocity, 
and $R_{\rm d}$ is the radius of the donor.

Finally, the mass loss rate of the donor, combining the mass transfer rate and the mass stripping rate, is
\begin{equation}
    \dot{M}_{\rm d} = \dot{M}_{\rm t} + \dot{M}_{\rm str}. 
\end{equation}
where the mass transfer rate considering the mass stripping effect is given by Equations (27)-(29) in \cite{1999ApJ...522..487H}.
Then the WD growth rate 
is
\begin{equation}
    \dot{M}_{\rm WD} = \eta_{\rm H} \, \eta_{\rm He} \, |\dot{M}_{\rm t}| = \eta_{\rm H} \, \eta_{\rm He} \, |\dot{M}_{\rm d}-\dot{M}_{\rm str}|, 
\end{equation}
and the wind mass loss rate is given by
\begin{equation}
    \dot{M}_{\rm wind} = \dot{M}_{\rm t} - \dot{M}_{\rm WD}.
\end{equation}

In addition, the wind also carries some specific angular momentum and can be estimated as $l_{\rm wind} = \left( \frac{q}{1+q} \right)^2$ in units of $A^2 \Omega_{\rm orb}$, where $A$ is the binary separation, and $\Omega_{\rm orb}$ is the orbital angular velocity. This will contribute to the transfer of total angular momentum $J$ and
\begin{equation}
\left( \frac{\dot{J}}{\dot{M}}\right)_{\rm wind} = l_{\rm wind} A^2 \Omega_{\rm orb},
\end{equation}
where $M$ is the total mass in the system.

We implement the above described optically thick wind model and mass-stripping effect in \texttt{MESA} by modifying the binary test-suite models {\tt star\_plus\_point\_mass}.
The wind mass transfer rates and mass loss/growth rates are computed using the {\tt other\_adjust\_mdot} module, while the total angular momentum is adjusted in the {\tt other\_extra\_jdot} module. We also use the {\tt other\_rlo\_mdot} module to compute the mass transfer rate 
\citep{2015ApJS..220...15P}. 

Our binary stellar evolution simulations start from a WD and a zero-age MS star, whose initial chemical composition is set to $\rm (X, Y, Z) = (0.70, 0.28, 0.02)$.
Note that the WD is represented as a point mass; therefore, the effects of composition changes and nuclear burning on the accreting WD are not included in this study.
When the MS star evolves into the RG stage, its envelope significantly expands and eventually reaches the effective Roche-lobe radius \citep{1983ApJ...268..368E}, resulting in the start of RLOF process.
We set the RLOF to be turned on when the donor radius fulfills 99.9\% of the Roche-lobe radius.
When the mass transfer rate reaches the critical accretion rate, the optically thick wind model described in Equations~(1)-(9) is applied.
Until the mass transfer rate drops below the critical accretion rate again, the WD wind stops, but the mass transfer keeps until reaching one of our terminal criteria.

We terminate the binary evolution based on five different terminal criteria: 
(1)~If the WD reaches the Chandrasekhar mass $M_{\rm Ch}=1.38 \, \rm M_{\odot}$, the model is denoted as a successful SN~Ia progenitor system. 
(2)~If the mass transfer rate becomes lower than $\dot{M}_{\rm low}$, the H-shell burning of the donor becomes unstable, and the model is denoted as occurring a H shell flash (or so-called a nova) \citep{1999ApJ...522..487H}. 
(3)~If the mass transfer cannot remain in a stable state at the beginning, the model is denoted as entering the common envelope phase \citep{1993PASP..105.1373I}.
(4)~If the donor's surface luminosity exceeds the local Eddington luminosity and the surface cannot be resolved by {\tt MESA} during the evolution, the model is denoted as facing the surface resolution limit.
Note that this criterion does not rule out potential SN~Ia progenitors; rather, it highlights that our {\tt MESA} simulations are unable to explore the viability of these models and may require additional physical assumptions regarding the stellar surface.
(5)~Lastly, if the donor radius does not reach the Roche-lobe radius or the RLOF ceases before the WD reaches $M_{\rm Ch}$, the model is denoted as no RLOF.
Depending on the initial configuration, the final outcome (when a SN~Ia occurs) of our donor star models may reach the RG phase with or without prior helium flash, or the AGB phase with or without thermal pulsing. 
The classification of these types of companions will be described in detail in Section~\ref{sec: BE_final_outcome}.

\subsection{Post-impact Evolution of the Evolved Companion}

The interaction between SN ejecta and an RG companion star in a Type Ia supernova has been investigated in our previous work through multi-dimensional hydrodynamic simulations \citep{2012ApJ...750..151P}.
The simulations included an RG companion with a mass of 0.98 $\rm M_\odot$, a W7-like explosion model, and three sets of binary separations ranging from approximately 3 to 5 times the RG radius.
The results indicate that, in all cases, more than 95\% of the RG envelope is stripped by the SN ejecta due to the envelope's lower pressure compared to the ram pressure of the ejecta.
In addition, the dependence of envelope mass stripping on binary separation follows a power-law relation. Earlier studies, such as \cite{1992ApJ...399..665L} and \cite{2000ApJS..128..615M}, also concluded that RG companions typically lose nearly their entire envelopes following SN impacts.

However, earlier hydrodynamic simulations could not resolve the stellar core of a giant star or the H and He burning shells.
Given that most of our RG and AGB models have similar envelope structures and binary separation-to-radius ratios, we adopt this conclusion for these models as well. However, this conclusion does not apply to our post-He-flash RG companions, as their binary separation-to-donor radius ratios ($A/R_{\rm d}$) at the time of explosion are significantly larger, making the SN impact relatively minor. We discuss the post-He-flash RG models separately in Section~\ref{sec:post-He-flash_RG}. 
Additionally, previous hydrodynamic simulations did not fully incorporate certain aspects of microphysics, such as detailed treatments of radiative transfer, radiative cooling, and nuclear burning within the stellar envelope. These physical processes might influence the precise fraction of envelope material that survives after the SN impact. Therefore, the exact percentage of the envelope that remains uncertain. In this study, guided by previous numerical results (e.g. \cite{2012ApJ...750..151P}), we conservatively assume that a small fraction ($<5\%$) of the envelope may persist after the impact.

To search for possible surviving companions in Type Ia supernova remnants, it is necessary to simulate the evolution of a surviving companion for thousands to tens of thousands of years after a SN explosion. However, this timescale cannot be achieved with hydrodynamical simulations because the time step (on the order of seconds or minutes) is constrained by the local sound speed. Therefore, we use {\tt MESA} to conduct the long-term post-impact evolution. We approximate the dynamical effects of SN-companion interactions with two one-dimensional effects: (1) mass-stripping from the companion's envelope, and (2) the SN heating effect due to energy deposition. This type of toy model has been widely used in previous studies to search for surviving companions through various progenitor channels \citep{2012ApJ...760...21P,2013ApJ...773...49P, 2019ApJ...887...68B, 2022ApJ...933...38R,2022ApJ...928..146L,2023ApJ...949..121C,2024ApJ...973...65W}.
In addition, we investigate the post-impact evolution of these loose-envelope surviving companions using a toy model with two scenarios: one where only a stellar core remains (hereafter referred to as the ``core-only scenario") and the other where a small portion of the envelope also remains (hereafter referred to as the ``remaining-envelope scenario").

We perform post-impact evolution simulations of the surviving companion in {\tt MESA}. 
The simulation steps in our toy model are as follows: 
First, we strip the envelope of a companion star by using the {\tt relax\_mass} module in {\tt MESA} to mimic the envelope-stripping effect during the SN-companion interaction. The core-only models involve complete stripping of the envelope (with a surface composition identical to that of the core), whereas the remaining-envelope models retain a small portion of the envelope, which will be discussed in detail in Section~\ref{sec:result_RET}. The maximum mass loss rate is fixed to $10^{-5}$ $\rm M_{\odot}$~yr$^{-1}$ ({\tt lg\_max\_abs\_mdot=-5}) for the stability of the relaxation, because higher mass-loss rates result in unstable mass loss, while smaller rates are less consistent with the rapid stripping observed in the hydrodynamics simulations \citep{2012ApJ...750..151P}.
We assume that the companion star returns to hydrostatic equilibrium immediately after the SN~Ia explosion, as the dynamical timescale of the SN-companion interaction (~days) is much shorter than the thermal timescale of the post-impact evolution ($\sim 10^3$ yr). While our toy model does not simulate the hydrodynamic impact phase, the post-SN structure is initialized using a stripped and heated hydrostatic model in MESA, allowing us to follow its long-term thermal relaxation.

Once the envelope is stripped, the companion is heated by the energy injection from the SN ejecta.
The total injected energy for core-only companion models is calculated by the cross section of the remaining core,
\begin{equation}
    E_{\rm inj} = \frac{f_{\rm abs}}{4}\left(\frac{R_{\rm core}}{A_{\rm f}}\right)^2 × E_{\rm SN},
    \label{eq_Einj}
\end{equation}
where $R_{\rm core}$ is the companion's core radius, $f_{\rm abs}=0.8$ is the assumed absorption fraction, $E_{\rm SN}= 1.233 \times 10^{51}$~erg is the supernova energy \citep{1984ApJ...286..644N, 2012ApJ...760...21P}, and $A_{\rm f}$ is the binary separation at the onset of SN~Ia event.

Once the amount of SN heating is determined, a heating profile based on the normalized mass coordinate $m(r)$ is applied.
The profile consists of an uniform heating distribution from the surface to a depth $D_{\rm h}$ in mass coordinates, transitioning to a half-normal distribution with a width $\sigma$ for regions deeper than $D_{\rm h}$.
The heating profile can be expressed by
\begin{equation}
\epsilon(m) = \begin{cases}
                \epsilon_0 \exp^{-(\frac{m(r)-(1-D_{\rm h})}{2\sigma})^2}, & {\rm if }~m(r) < 1-D_{\rm h}\\
                \epsilon_0, & {\rm otherwise}
               \end{cases} \label{eq_heat}
\end{equation}
where $\sigma=0.01$, $D_{\rm h}$ = 0.05, and
$\epsilon_0$ is a heating constant that ensures the total SN heating over a heating timescale $t_{\rm h} = 10 \, \text{days}$ will be equal to $E_{\rm inj}$. Note that we have tested the results using different values of $D_{\rm h}$ and $E_{\rm inj}$. We find that the post-impact evolution is dominated by the thermal relaxation following the sudden envelope stripping, rather than the detailed structure of the heating. Therefore, the exact values of this heating profiles has minor impact on our lateron evolution.

For the remaining-envelope models, since the companion radius is highly sensitive to the amount of remaining envelope and the inflation of the envelope is mainly due to the SN heating and shell burning, we evaluate $E_{\rm inj}$ based on the core radius instead of the companion radius. However, the heating profile required additional consideration. 
$\epsilon_0$ is first calculated based on the stellar core radius using Equation~\ref{eq_heat} and is then extended uniformly to the envelope portion. This approach results in a total injected energy that is slightly higher than the value estimated from Equation~\ref{eq_Einj}.
Finally, the surviving companion evolves until reaching the simulation time $t_{\rm max}=10^5$ yr, which is set to be $t_{\rm pe} \sim 10^5$ yr based on the thermal timescale of SN impact (e.g. \cite{2012ApJ...750..151P}). Furthermore, a supernova remnant will hardly be recognized if it is older than $10^6$ years.

\section{RESULTS}
\label{sec:res}

In this section, we present a series of simulations of the symbiotic binaries leading to SNe~Ia. 
In Section~\ref{sec:result_BE}, by implementing \cite{1999ApJ...522..487H}'s optically thick wind model, we showcase the outcomes of wide parameter space of binary evolution simulations and demonstrate four types of companion stars (donor) in the SN~Ia progenitor systems. 
Following in Section~\ref{sec:result_postSN}, based on the results of previous hydrodynamic simulations of SN-companion interactions, which indicated that the donor envelope is almost completely stripped by the SN ejecta, we present the post-SN stellar evolution simulations of two scenarios: models of the companion's surviving stellar core (core-only scenario), comparing them to additional models considering a small layer of remaining-envelope material (remaining-envelope scenario). 
Lastly, in Section~\ref{sec:obs}, we explore the observability of the surviving companion in the two scenarios. 

\begin{figure}
\epsscale{1.2}
\plotone{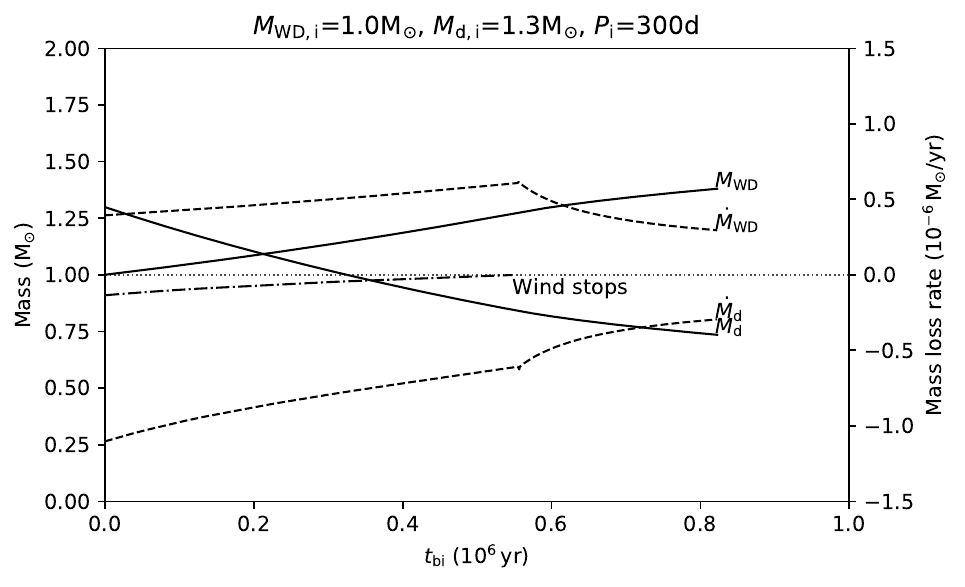}
\caption{
Time evolution of a SN~Ia progenitor model with an initial WD mass $M_{\rm WD, i} = 1.0 \rm M_\odot$ and an initial donor mass $M_{\rm d,i} = 1.3 \rm M_\odot$ of a red giant donor. The initial orbital period is set to $P_{\rm i}=300 \,\rm day$.
$t_{\rm bi} = 0 \rm \, yr$ marks the beginning of binary interaction. The solid lines represent the masses of the accreting WD ($M_{\rm WD}$) and the donor ($M_{\rm d}$).
The dashed lines show the corresponding mass loss rates, and the dash-dotted line indicates the WD wind mass loss rate ($\dot{M}_{\rm wind}$).
}
\label{fig:hachisu_fig10}
\end{figure}

\begin{figure*}
\epsscale{1.15}
\plotone{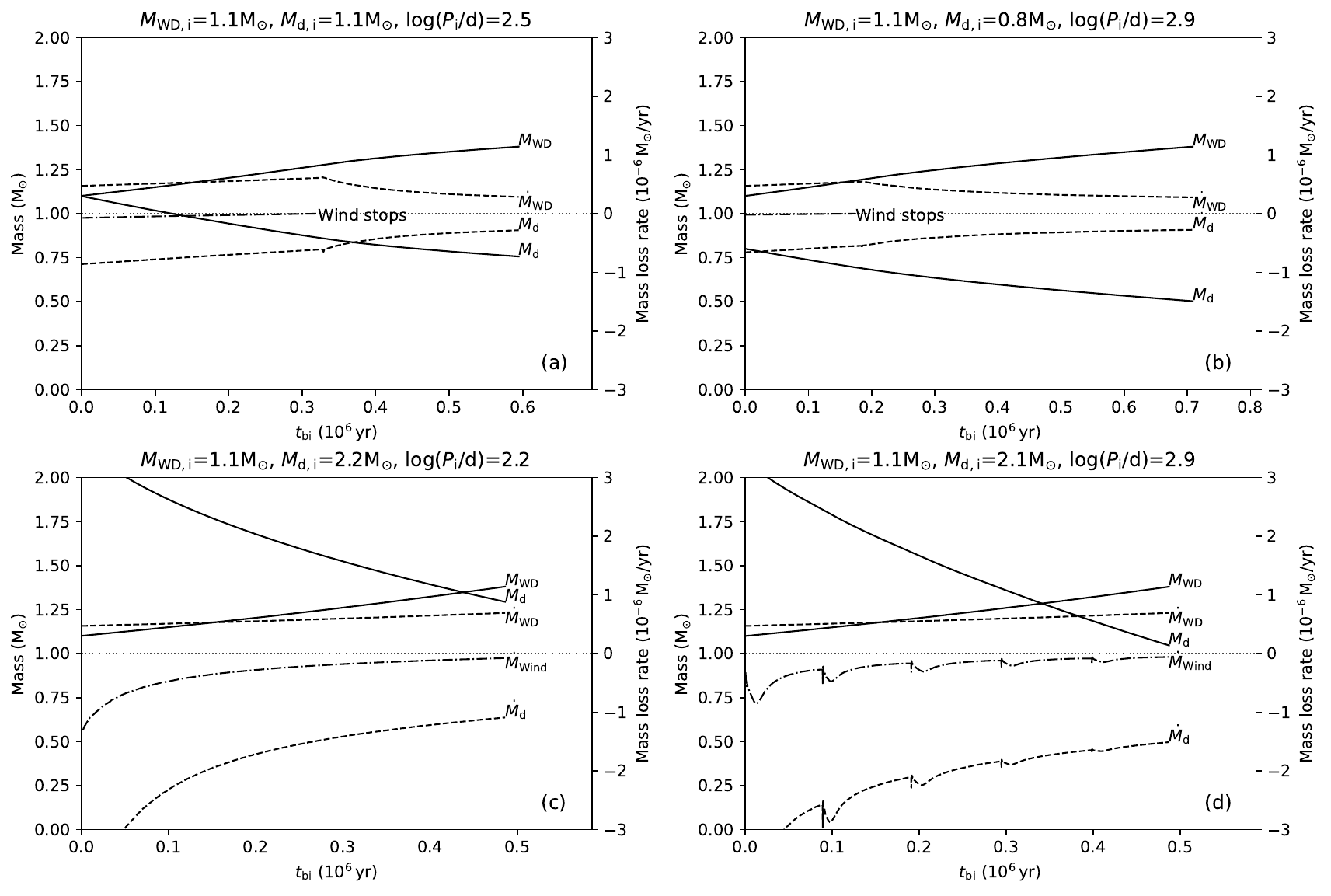}
\caption{Same as Figure~\ref{fig:hachisu_fig10} but for four SN progenitor models with initial WD masses $M_{\rm WD, i}=1.1 \, \rm M_{\odot}$ but different initial orbital periods and donor masses.. Different panels represent different typical donor types of SN~Ia progenitors in the binary evolution: (a) a red giant, (b) a post-He-flash red giant, (c) an AGB star, and (d) a TP-AGB star. The identification of these four final types of the donors is introduced in Section~\ref{sec: BE_final_outcome}.}
\label{fig:BE_tracks}
\end{figure*}

\begin{figure*}
\epsscale{1.2}
\plotone{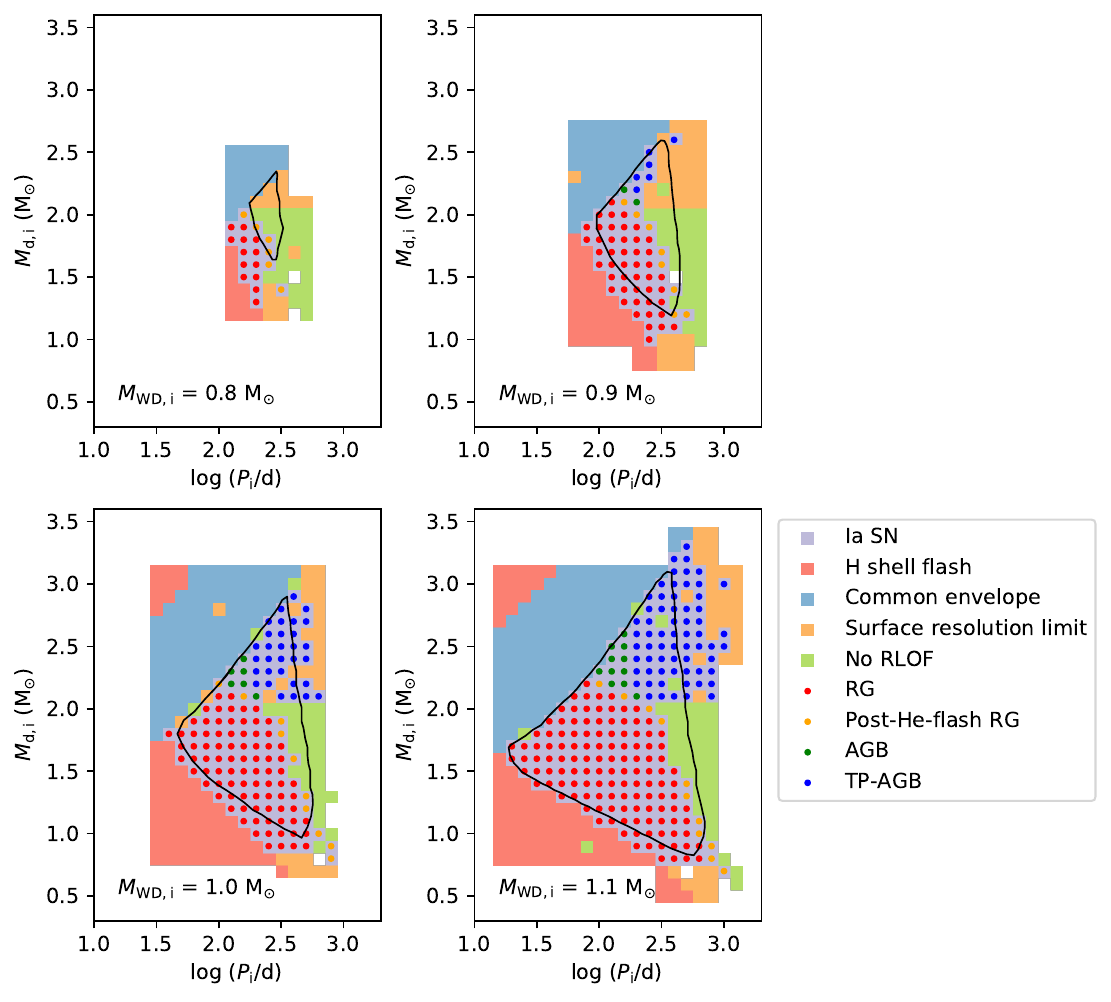}
\caption{The initial condition distribution with binary evolution outcomes. Four panels vary in the initial WD mass, $M_{\rm WD, \,i}=0.8, \,0.9, \,1.0, \, 1.1 \, \rm M_{\odot}$, respectively. Each panel shows the parameter space of $\log$-scaled initial binary period $\log P_{\rm i}$ (x-axis) and initial donor mass $M_{\rm d, \,i}$ (y-axis). The grid size is 0.1 $\log$ day on the x-axis and 0.1 $\rm M_{\odot}$ on the y-axis. Grid colors represent the outcome types: SN~Ia progenitor (purple), H shell flash (red), common envelope (blue), surface resolution limit (orange), and no RLOF (green). The colored dots upon those SN~Ia progenitor models represent the final conditions of donors in SN~Ia progenitor systems: the RG (red dot), the post-He-flash RG (orange dot), the AGB (green dot), and the TP-AGB (blue dot) stages. The black contours are the regions of SN progenitors proposed by \cite{1999ApJ...522..487H} (their Figures~11 and 12).}
\label{fig:binary_result}
\end{figure*}

\subsection{SNe~Ia Progenitors}
\label{sec:result_BE}

\subsubsection{Code Validation}

To validate our implementation of the optically thick wind model in {\tt MESA}, we initially attempt to replicate a standard binary evolution model outlined in Figure 10 in \cite{1999ApJ...522..487H}, which has an initial WD mass $M_{\rm WD,i} = 1.0 \rm M_\odot$, a donor star mass $M_{\rm d,i} = 1.3 \rm M_\odot$, and binary orbital periods $P_{\rm i}=300\,\rm day$.
The binary evolution starts with an MS companion and a point-mass WD. 
When the companion evolves into the red giant stage and fills its Roche lobe, the binary interaction starts, and the optically thick wind starts blowing.
Figure~\ref{fig:hachisu_fig10} demonstrates the evolution of the masses ($M_{\rm d}$ and $M_{\rm WD}$) and the mass change rates ($\dot{M}_{\rm wind}$, $\dot{M}_{\rm d}$ and $\dot{M}_{\rm WD}$) of the binary starting, starting from the onset of the binary interaction, defined as  $t_{\rm bi}=0$ yr.
The optically thick wind starts with an initial wind rate $\dot{M}_{\rm wind}=-3.94\times10^{-7} \, \rm M_{\odot} \, yr^{-1}$ and further induces the mass stripping on the donor. Then the WD wind gradually decreases due to the decreasing mass ratio $q$, until the wind rate becomes lower than the critical rate, $\dot{M}_{\rm wind} < \dot{M}_{\rm cr}$, stopping the wind at $t_{\rm bi}=5.5 \times 10^{5} \rm yr$. After the wind stops, the WD keeps growing until it reaches $M_{\rm Ch}$ and explodes at $t_{\rm bi}=8.1 \times 10^{5}\rm yr$. The total delay time of the SN explosion since the beginning of the binary system is $\tau$=4.6 Gyr and the final donor mass is $M_{\rm d, f}=$0.73 $\rm M_{\odot}$. 

Compared to Figure~10 of \cite{1999ApJ...522..487H}, the overall evolutionary tracks and the final donor mass in our model show a high level of consistency, albeit evolving slightly faster. This difference is likely due to \cite{1999ApJ...522..487H} using the empirical formula from \cite{1983ApJ...270..678W} in their calculations, as well as differences in the microphysics treatments in {\tt MESA}. We also note that the values of $\eta_{\rm He}$ calculated in our simulations are somewhat higher than the value adopted in \cite{1999ApJ...522..487H}, which may help explain the slight discrepancies between our results and theirs. A higher $\eta_{\rm He}$ in our simulations could enhance He-shell burning on the WD surface, leading to a modest increase in the WD mass and, consequently, a difference in the delay time compared to the values presented in \cite{1999ApJ...522..487H}.
We have also checked other models in our considered parameter space, and in the later subsections, we will show a comparison of our final outcomes with \cite{1999ApJ...522..487H}'s results. 
Therefore, we affirm that we correctly apply the optically thick wind model from \cite{1999ApJ...522..487H} in {\tt MESA} binary evolution simulations. 

\subsubsection{Evolutionary Tracks toward SNe~Ia}
\label{sec:evo_curves}

We apply the optically thick wind model to perform the binary evolution simulations. The parameter space, which is taken to match the parameters used in \cite{1999ApJ...522..487H}, ranges within $M_{\rm WD, i}=0.8, 0.9, 1.0, 1.1 \,\rm M_{\odot}$, $M_{\rm d, i}= 0.5-3.5 \,\rm M_{\odot}$ with an interval of $0.1 \,\rm M_{\odot}$, and $P_{\rm i}= 10^{1.1-3.1} \,\rm day$ with an interval of $10^{0.1} \,\rm day$. In total, we perform 1,260 {\tt MESA} binary evolution simulations. 

Figure~\ref{fig:BE_tracks} illustrates four typical evolutionary tracks with an initial WD mass $M_{\rm WD,i}=1.1 \,\rm M_\odot$ but having different initial donor masses $M_{\rm d,i}$ or orbital periods $P_i$. 
Panels~(a)~and~(b) represent models with smaller initial donor masses, while panels~(c)~and~(d) feature larger initial masses. Among models with similar initial donor masses, those in panels~(b)~and~(d) exhibit larger binary separations than models in panels~(a)~and~(c). 
In the upper panels, the WD winds stop during the mass transfer, and their WDs become SNe~Ia at $t_{\rm bi}=5.95 \times 10^5\,\rm yr$ ($\tau$=8.6 Gyr) in panel~(a) and $t_{\rm bi}=7.09 \times 10^5$ yr ($\tau$=27.9 Gyr) in panel~(b). 
Despite having similar evolution processes, the donors in panels (a) and (b) will evolve into different evolutionary stages at the onset of the explosion due to different evolution timescales. We will further describe the final outcomes in Section~\ref{sec: BE_final_outcome}.
On the other hand, more massive donors evolve more rapidly until their WDs become SNe~Ia at $t_{\rm bi}=4.86 \times 10^5$ yr ($\tau$=970 Myr) in panel~(c) and $t_{\rm bi}=4.87 \times 10^5$ yr ($\tau$=1.1 Gyr) in panel~(d). Their optically thick winds persist until the SN events because their mass ratios $q$ stay higher than $0.79$. 
In particular, panel~(d) clearly demonstrates the donor with a periodically varying mass loss rate $\dot{M}_{\rm d}$ due to He shell flashes, inducing pulsations of the WD wind. This feature was not captured by \cite{1999ApJ...522..487H} likely because they use the empirical formulae from \cite{1983ApJ...270..678W} to perform the binary evolution.
We find that the pulsing donor's mass loss rate is produced by a companion evolving to the Thermal-Pulsing AGB (TP-AGB) stage. In fact, both models in panels~(c)~and~(d) evolve into the AGB stage, which can be further identified by their interior structure and will be demonstrated in Section~\ref{sec: BE_final_outcome}.

\subsubsection{Binary Populations and Parameter Space}
\label{sec: BE_final_outcome}

Figure~\ref{fig:binary_result} displays the final outcomes of our binary evolution models. 
Each panel demonstrates the binary population with a fixed value of $M_{\rm WD, i}$ and varying $\log P_{\rm i}$ in the x-axis and $M_{\rm d, i}$ in the y-axis. The parameter space was described in Section~\ref{sec:evo_curves}.
Each parameter set is represented by a filled square as the grid, with its color indicating the final outcome based on their termination criteria (see Section~\ref{sec:method_BE}): purple for a SN~Ia progenitor system, red for ending with an H shell flash, blue for a common envelope, orange for encountering the numerical resolution limit on the donor's surface, or green for facing the cessation of RLOF.
Upon each purple grid representing a SN~Ia progenitor system, a colored dot further marks the final type of the donor: a RG star (red), a post-He-flash RG star (orange), an AGB star (green), or a TP-AGB star (blue).

Here, our focus lies on the overall distribution of the final outcomes represented by the grid colors in Figure~\ref{fig:binary_result}. 
To facilitate comparison, we delineate the SN~Ia region from \cite{1999ApJ...522..487H} in each panel using a black contour.
The distribution of models encountering common envelopes and H shell flahses (novae) aligns with \cite{1999ApJ...522..487H}'s results. 
However, some of our models with large initial orbital periods terminate because of the dramatically expanding donor during RLOF, resulting in negative luminosity values on the surface of the donor. These models were previously attributed to encountering He flashes or H shell flashes in \cite{1999ApJ...522..487H}.

Discrepancies emerge notably in the distribution of SN~Ia progenitor systems between our results and the previous study.
Panels in Figure~\ref{fig:binary_result} with smaller initial WD masses exhibit larger offsets in SN~Ia regions, a divergence attributed to conducting longer self-consistent binary evolution simulations in {\tt MESA}.
Models with smaller initial WD masses require longer evolution times to reach $M_{\rm ch}$, leading to larger differences that magnify deviations from earlier simulation codes.
Furthermore, the distributions of our SN~Ia progenitor systems in panels with $M_{\rm WD, i}=0.9, 1.0, \, 1.1 \, \rm M_{\odot}$ deviate from the simple triangular shapes of \cite{1999ApJ...522..487H}'s regions, displaying a concavity at the right border.
This discrepancy emerges from different conclusions regarding the final outcomes beyond the right border of the SN~Ia progenitor systems, where \cite{1999ApJ...522..487H} labeled the final outcomes as He flashes on the donors, deeming them incapable of producing SN~Ia explosions. 
However, our findings indicate that only models with initial donor masses $M_{\rm d, i} \lesssim$ 2.0 $\rm M_{\odot}$ and sufficiently large orbital periods may lead to this consequence (denoted as ``no RLOF" in Figure~\ref{fig:binary_result}), and most models with larger donor masses evolve into AGB or TP-AGB stars, contributing to the main discrepancy between \cite{1999ApJ...522..487H} and our results. 

\begin{figure}
\epsscale{1.1}
\plotone{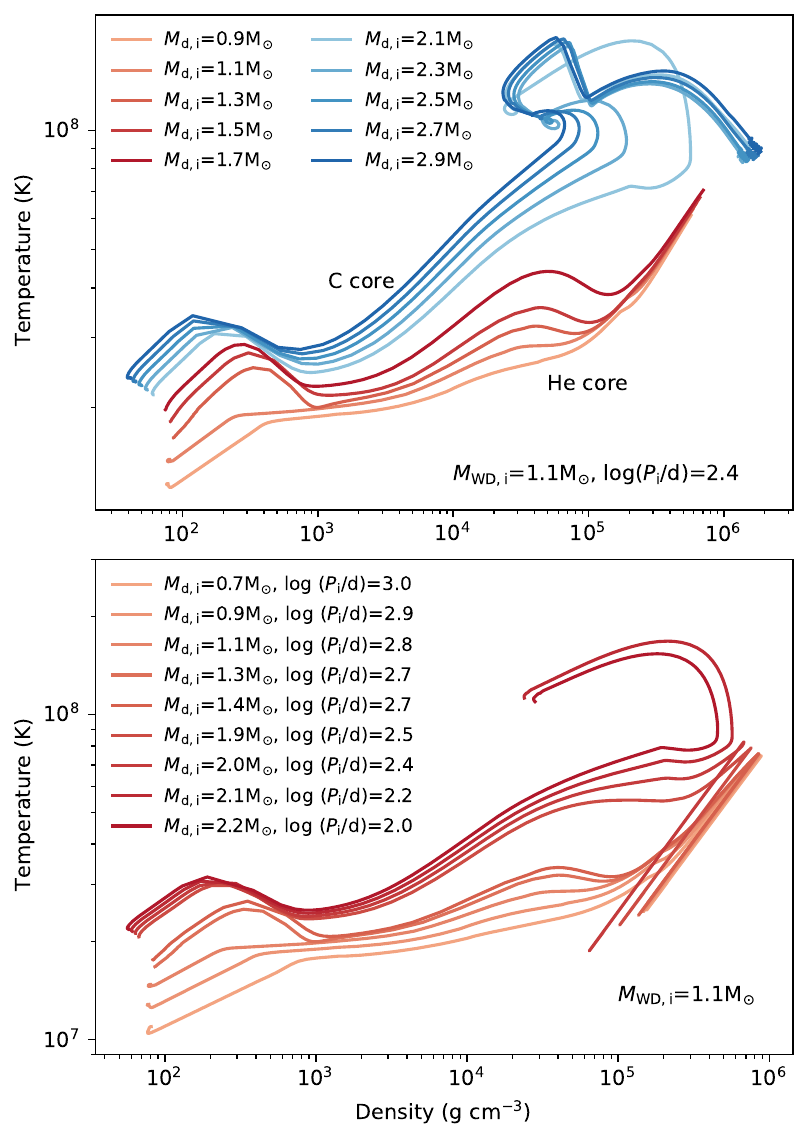}
\caption{Interior structure of the donors at the time of the SN explosion. Upper panel: The initial white dwarf mass and initial period are fixed; only the initial red giant mass varies. Redder colors are He core red giants, and bluer colors are C core AGB stars. Lower panel: Same as the upper panel but with post-He-flash red giants.}
\label{fig:rg_structure}
\end{figure}

\begin{figure*}
\epsscale{1.1}
\plotone{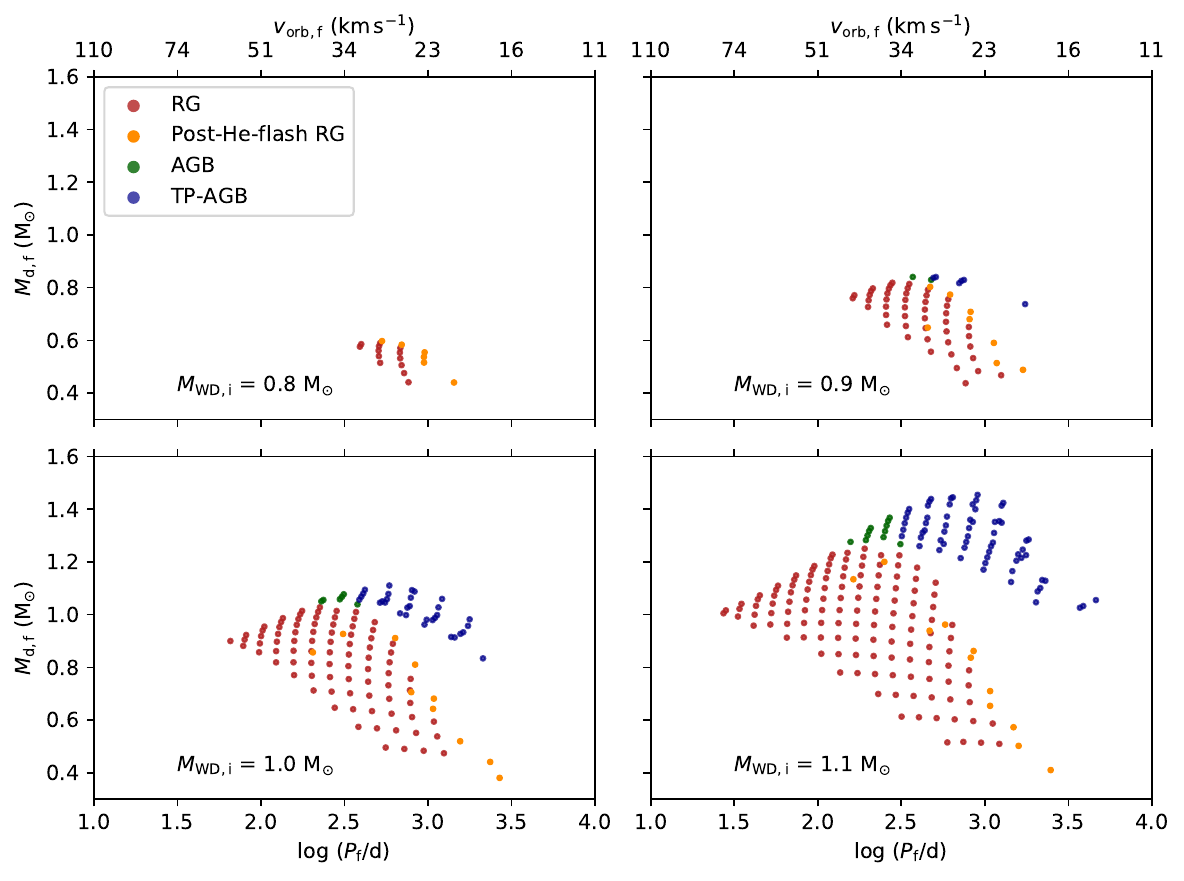}
\caption{The final conditions of SN~Ia progenitor systems. Four panels vary in the initial white dwarf mass, $M_{\rm WD, \,i}=0.8, \,0.9, \,1.0, \, 1.1 \, \rm M_{\odot}$, respectively. The final white dwarf mass is set to be $M_{\rm Ch}=1.38 \, \rm M_{\odot}$. Each dot shows the final binary period $P_{\rm f}$, the final orbital speed of the donor $v_{\rm orb,f}$, and the final companion mass $M_{\rm d, f}$ of one progenitor system. Different colors represent the final evolutionary stages of the companion at the time of SN explosion. The companions are classified into four groups: helium-core RGs (red dots), post-He-flash RGs (orange dots), carbon-core AGBs (green dots), and carbon-core TP-AGBs (blue dots).
}
\label{fig:binary_result_final}
\end{figure*}

\begin{figure}
\epsscale{1.1}
\plotone{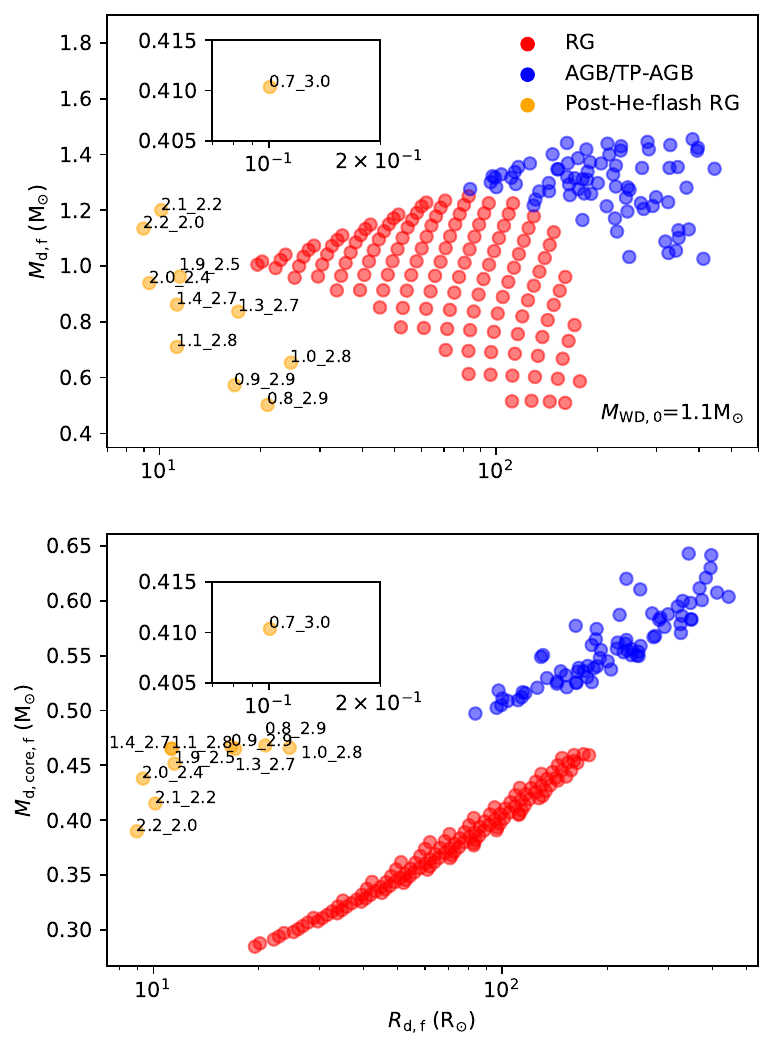}
\caption{The final mass (upper panel) and final core mass (lower panel) versus the radius of the companions in SN~Ia progenitor systems, where the initial white dwarf mass is 1.1 $\rm M_{\odot}$. Models with low radius-to-mass ratios are labeled with the initial companion mass in $\rm M_{\odot}$ (first number) and the initial binary period in $\rm \log day$ (second number). The colored dots follow the same definition as in Figure~\ref{fig:binary_result}, indicating the final types of the donors.}
\label{fig:M_logR}
\end{figure}

Previously mentioned in Section~\ref{sec:evo_curves}, TP-AGB companions are revealed with their pulsing mass loss rate.
The final types of donors can be further characterized through their interior structure. 
Figure~\ref{fig:rg_structure} compares the interior structure of different donor models at the time of SN explosion. The upper panel shows a set of models with varying initial donor masses while maintaining constant initial WD masses and orbital periods. 
The RG donors (red lines) lack sufficient central pressure for He flash or burning, while the AGB/TP-AGB donors (blue lines) successfully initiate He burning and form a central carbon (C) core.
The lower panel demonstrates the interior structure of the post-He-flash RGs:
donors with initial masses $M_{\rm d, i} \lesssim$ 2.0 $\rm M_{\odot}$ exhibit consistent electron degeneracy in the core's innermost region, indicating that they underwent an off-center He flash \cite[][]{2011ApJS..192....3P}.
Conversely, donors with initial masses $M_{\rm d, i}>$ 2.0 $\rm M_{\odot}$ (the uppermost two lines) follow a distinct trajectory as they ignite He at a lower degree of degeneracy \cite[][]{2011ApJS..192....3P}, resulting in a small fraction ($<5\%$) of carbon forming in the core, while the core still primarily consisting of helium.

As previously stated in the preceding paragraph, in \cite{1999ApJ...522..487H}'s study, models with donors that have experienced He flashes are denoted as unsuccessful in forming SNe~Ia and distribute beyond the right border of the SN~Ia regions. However, our results indicate that only models with initial donor masses $M_{\rm d, i} \lesssim$ 2.0 $\rm M_{\odot}$ may lead to He flashes (with a very few exceptions that have initial donor masses between $2.0 < M_{\rm d, i} < 2.2 \,\rm M_{\odot}$); models with larger initial donor masses are able to form AGB/TP-AGB donors without facing the unstable mass transfer. 
Therefore, a concavity forms at the right border of the SN~Ia region with an upper boundary at $M_{\rm d, i}$ =2.0 $\rm M_{\odot}$ in each panel of Figure~\ref{fig:binary_result}. 

Figure~\ref{fig:binary_result_final} shows the final condition distribution of the SN~Ia progenitor systems, plotting final orbital period $P_{\rm f}$ and corresponding orbital speed $v_{\rm orb, f}$ on the x-axis, and final donor mass $M_{\rm d, f}$ on the y-axis.
The final orbital speeds for the donors are ranging from 21 to 32 km $\rm s^{-1}$, 20 to 44 km $\rm s^{-1}$, 17 to 59 km $\rm s^{-1}$, and 14 to 79 km $\rm s^{-1}$ corresponding to models with the initial WD masses $M_{\rm WD, i}=$ 0.8, 0.9, 1.0, and 1.1 $\rm M_{\odot}$.
Dots with different colors, following the same definition in the initial condition distribution in Figure~\ref{fig:binary_result}, label the final donor types corresponding to each set of the final conditions. 
Compared to the initial conditions, the final condition distribution of most of the models remains in order, except the post-He-flash RGs (orange dots in Figure~\ref{fig:binary_result} and \ref{fig:binary_result_final}).
This is because most of the SN~Ia progenitor models go through stable RLOF until their WDs become SNe. 
On the contrary, models that have experienced He flashes lose their stability in mass transfer after the flashes. 
This feature can be more easily identified in the upper panel of Figure~\ref{fig:M_logR}, which shows the same models with $M_{\rm WD, i}=1.1 \, \rm M_{\odot}$ as in Figure~\ref{fig:binary_result}, but with the x-axis replaced by the final donor radius, demonstrating a significant drop in radius for donors underwent He flashes. This radius-dropping effect causes the cessation of mass transfer in most of the cases, forming the models with ``no RLOF" outcome in Figure~\ref{fig:binary_result}. The very few successful SN~Ia progenitor models form at the time point when a He flash happens right before the WD becomes a SN. Moreover, these models serve as the boundary between the two donor core types in SN~Ia progenitor systems in the initial condition map (Figure~\ref{fig:binary_result}): above the boundary are C core AGB/TP-AGB stars; on and below the boundary are RG stars.

Now, we have demonstrated the four types of SN~Ia progenitor models with different final donor types identified through various aspects. 
Here, we briefly summarize the features of these four types of models:

\begin{enumerate}
\item Models with lower initial donor masses and orbital periods have the final donors staying in the RG stage, same as the companions in \cite{1999ApJ...522..487H}'s SN~Ia progenitor models. Their WD winds stop during the RLOF, and stable mass transfer continues until their SNe occur.

\item A small portion of models have donors in the RG stage that have experienced He flashes and become He-rich; the donor's innermost interior structure reveals the occurrence of a He flash. Their WD winds stop during RLOF, and stable mass transfer continues until the He flashes take place, and the WDs become SNe soon after that.

\item Models with larger initial donor masses and orbital periods lead to C core formation in the donors, evolving into AGB stars. They can be identified by their interior structure and composition. Their WD winds persist during RLOF until the WDs become SNe~Ia. Stable mass transfer keeps until the SN occurs.

\item Most of the AGB companions evolve into the TP-AGB stage in our SN~Ia progenitor systems. They can be identified by their final interior structure, composition, and evolutionary curves. Their WD winds persist during RLOF until the WDs become SNe~Ia. Stable mass transfer keeps until the SN occurs.
\end{enumerate}

\begin{figure*}
\epsscale{1.2}
\plotone{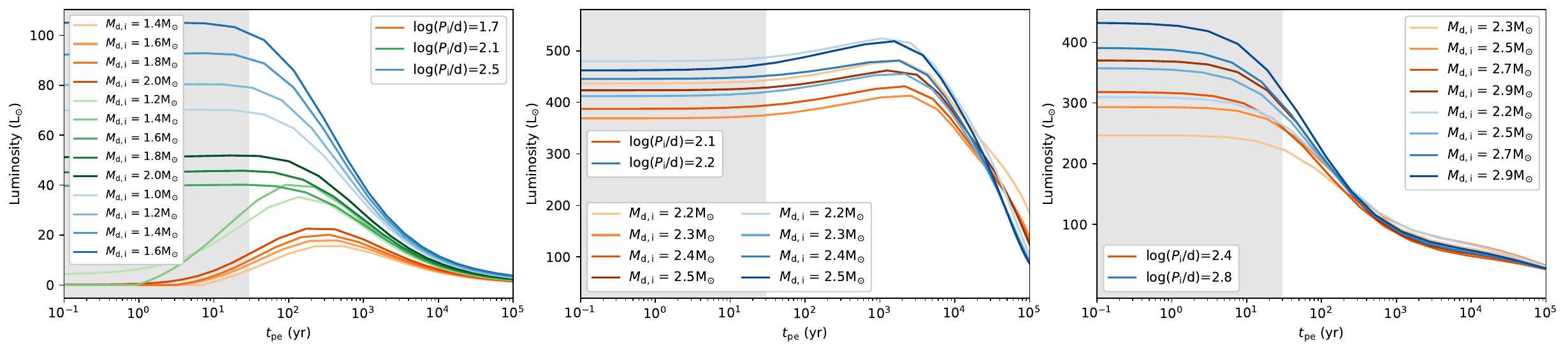}
\caption{Post-impact evolution of bolometric luminosity for the surviving stellar core models, whose initial WD masses $M_{\rm WD, i}=1.1 \, \rm M_{\odot}$. Different panels show the three different types of companions classified in Section \ref{sec: BE_final_outcome}: RG (left), AGB (middle), and TP-AGB (right) models. $t_{\rm pe} = 0 \, \rm yr$ marks the occurrence of the SN, and the gray area indicates the time period within the first 30 years following the SN explosion. This time window has greater uncertainty due to the mass loss procedure we employed. The heating timescale is set at 3 days.}
\label{fig:L_t_coreonly}
\end{figure*}

\begin{figure}
\epsscale{1.1}
\plotone{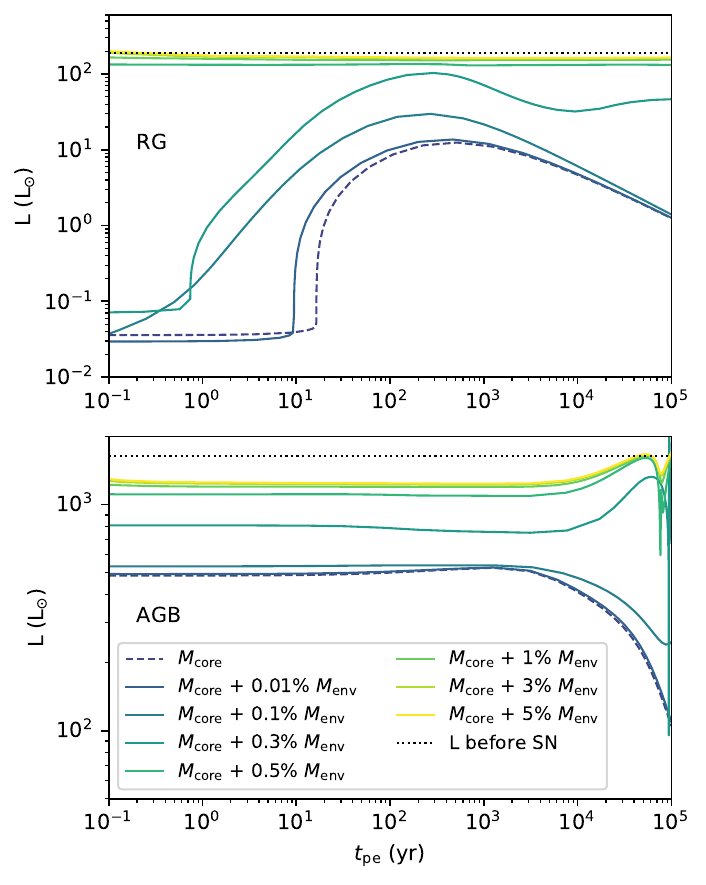}
\caption{Test of the remaining-envelope material in the post-impact evolution of companion models with $M_{\rm WD, i}=1.1 \,\rm M_{\odot}$. $t_{\rm pe} = 0 \, \rm yr$ marks the occurrence of the SN. Two models are depicted: $M_{\rm d, i}=1.8\,\rm M_{\odot}$, $P_{\rm i}=10^{1.5}\, \rm day$ with a He core (upper panel) and $M_{\rm d, i}=2.2\rm M_{\odot}$, $P_{\rm i}=10^{2.2}\, \rm day$ with a CO core (lower panel). In each panel, the dashed line represents the bolometric luminosity evolutionary curve of a core-only model, the dotted line represents the donor's luminosity prior to SN impact, and the other colored lines show the evolutionary curves of models with varying amounts of remaining envelope, indicated by the percentage of final envelope mass $M_{\rm env}$ in the end of binary evolution.}
\label{fig:RET}
\end{figure}


\subsection{Post-impact Evolution of the Surviving Companion}
\label{sec:result_postSN}

In this Section, we further investigate the heating effect during the post-impact evolution of the remaining stellar cores. Additionally, to ensure the robustness of our findings in a real-world scenario, we explore a situation considering a small portion of remaining envelope. 

\subsubsection{Core-only Scenario}
\label{sec:result_core-only}

We uniformly selected a total of 46 SN~Ia progenitor models within the parameter space of $M_{\rm WD,i}=1.1 \,\rm M_\odot$ (Figure~\ref{fig:binary_result}, lower right panel). 
These models were chosen from either at the boundary or within the SN~Ia progenitor region, with a $\sim0.2 \rm M_{\odot}$ interval for the initial donor mass $M_{\rm d, i}$ and a $10^{0.2}$ day interval for the orbital period $P_{\rm i}$.
Figure~\ref{fig:L_t_coreonly} shows the bolometric light curves of these core-only models, with different panels referring to various types of models identified in the previous step. 
Note that because the light curves are sensitive to the settings of the mass loss rate within our toy model, the evolutionary curves within the first 30 years may have relatively larger uncertainties (marked by a gray area in each panel).
The increase in bolometric luminosity is mainly due to the strong heating from the burning core, which inflates the envelope of the surviving companion. Depending on the model, the bolometric luminosity may peak before or after the heating timescale. Once it reaches equilibrium, it will eventually decrease as it releases gravitational energy due to core contraction.
The left panel shows the evolutionary tracks of RG core models. There are three sets of initial orbital periods, each varying initial donor masses with an interval of 0.2 $\rm M_{\odot}$. 
For models with the same orbital periods but different donor masses, larger initial donor masses result in larger final cores, making them brighter after the SN impact. For different sets of initial orbital periods, larger initial orbital periods lead to brighter remaining cores because they evolve into later stages prior to SN explosion. 
All RG core models exhibit much dimmer luminosity peaks than the values prior to the SN ($L \sim 250-2000 \,\rm L_{\odot}$).

The middle and the right panels display the AGB and TP-AGB core models, respectively. Their luminosities are much higher than the RG core models because of their higher central temperature but with similar core radius sizes. 
Among these models, AGB cores evolve slower than TP-AGB cores and are able to stay in the bright stage for a longer period. 
Both AGB and TP-AGB cores are much dimmer than the values prior to SN ($L \sim 1200-1700 \,\rm L_{\odot}$ for AGB stars, and $L \sim 2500-5700 \,\rm L_{\odot}$ for TP-AGB stars).
By the end of the evolution ($t_{\rm pe}=10^5 \, \rm yr$), all surviving donor cores evolve toward the WD branch. 
Overall, for evolved companions before the SN event, the surviving cores are typically fainter than their state prior to the SN but are mostly brighter than isolated cold WDs ($<10^{-2}$ $\rm L_{\odot}$). 
In addition, heating from the SN ejecta has almost no influence on any of the remaining donor cores.

\subsubsection{Remaining-envelope Scenario}
\label{sec:result_RET}

To investigate the influence of remaining envelope on the post-impact evolution of surviving companion, we initially considered two models, one for a RG and the other for an AGB star, with seven different fractions of remaining envelope mass ranging from 0.01\% to 5\% of the companion's envelope mass prior to the SN. $M_{\rm env}$ is 0.32 $\rm M_{\odot}$ for the RG model and 0.52 $\rm M_{\odot}$ for the AGB model.
The composition of the remaining-envelope material is the same as the companion's hydrogen envelope before SN explosion, (X, Y, Z) = (0.66, 0.32, 0.02). 
Figure~\ref{fig:RET} shows the test results through bolometric light curves. Solid lines in sequential colors represent the curves considering various amounts of remaining envelope, comparing with the core-only counterpart (dashed line) and the luminosity before the SN impact (dotted line; note that this represents a single value, shown as a horizontal line in the figure for easier comparison).
The evolutionary curves in both panels demonstrate the presence of two groups of evolutionary curves, one closer to the core-only counterpart and the other closer to the luminosity prior to the SN event. 
The difference between these groups is determined by the presence of remaining hydrogen envelope: even a small amount of remaining hydrogen envelope enables the donor to retain a brightness level similar to its pre-SN stage. 
In addition, the subsequent evolutionary process can continue with this thin residual layer of remaining-envelope material. For instance, the AGB model in the lower panel later evolves into the TP-AGB stage after $\sim 6 \times 10^4$ yrs.
Conversely, if the hydrogen envelope is completely stripped, the evolutionary curves tend to rapidly converge towards the curve of the core-only model, resulting in a potentially much fainter state than the donor's pre-SN impact stage.

With the test result, we generate additional 50 remaining-envelope models corresponding to the core-only models in Section~\ref{sec:result_core-only}.
Among all of our SN~Ia progenitor models, the donor's final conditions are primarily determined by the binary evolution process, resulting in different amounts of remaining H envelope for each model. 
Although the exact amount of remaining envelope is unknown, our results suggest the $1-3\%$ of remaining envelope might give significant contribution on bolometric luminosity evolution. 

\subsection{Observability of the Surviving Companion: Red Giants and AGB/TP-AGB Stars}
\label{sec:obs}

\begin{figure*}
\epsscale{1.2}
\plotone{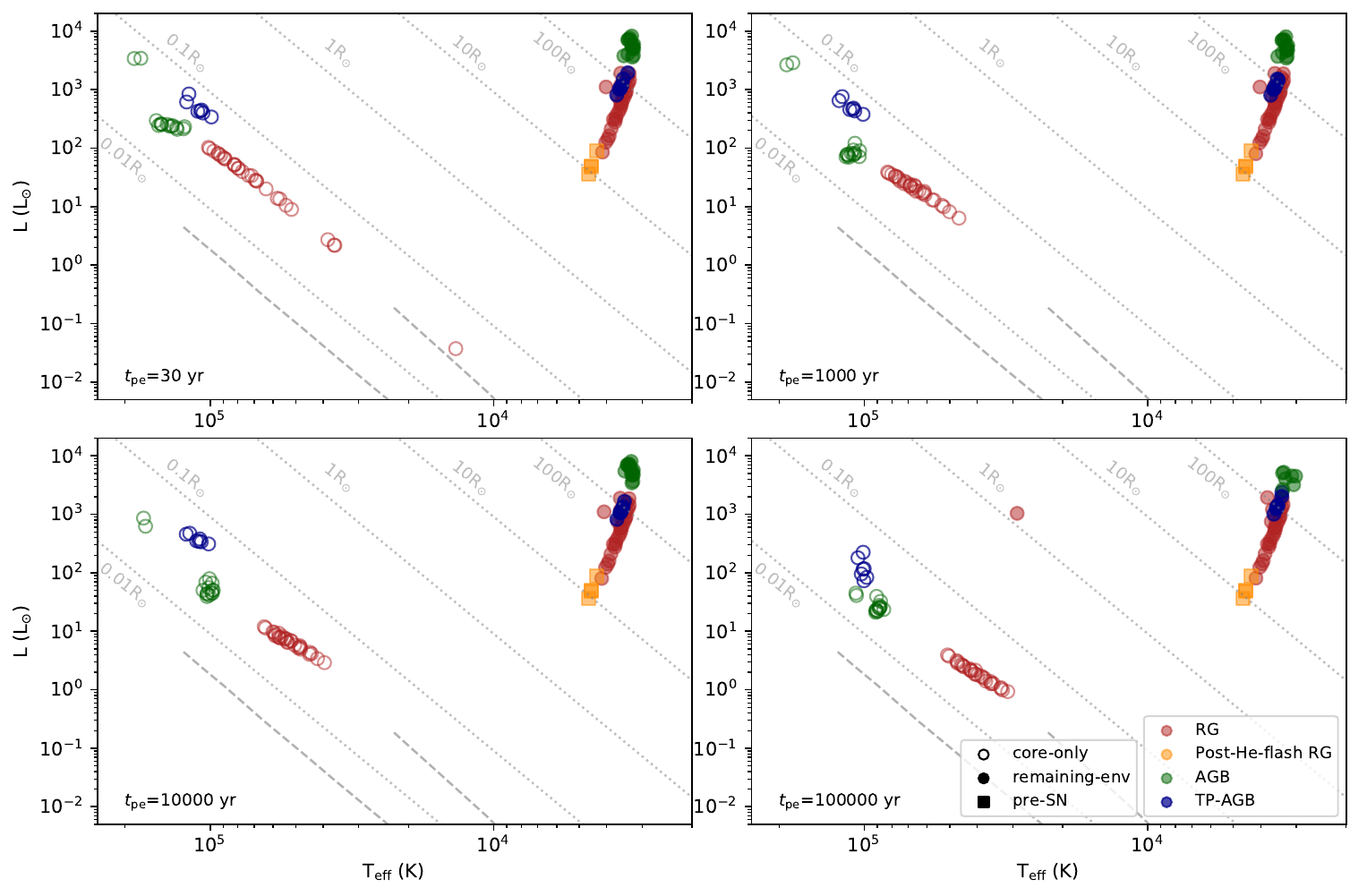}
\caption{H-R diagram with different panels varying post-SN evolutionary times ($t_{\rm pe}$). The core-only models are denoted by open circles, and the remaining-envelope models are represented by filled circles. Different colors are used to distinguish between the different donor types classified in the binary evolution simulations. Additionally, post-He-flash RG models are displayed with their pre-SN status, marked by squares. The gray dashed lines are the theoretical evolutionary tracks of 0.2 $\rm M_{\odot}$ (upper) and 1.3 $\rm M_{\odot}$ white dwarf models from \cite{2020ApJ...901...93B}.}
\label{fig:hr_diagram}
\end{figure*}

Combining the core-only models and their corresponding remaining-envelope counterparts with a $3\%$ remaining envelope, Figure~\ref{fig:hr_diagram} presents the post-impact evolution of these models on the Hertzsprung–Russell (H-R) diagram. 
Each panel varies the time after the SN event. The core-only models and remaining-envelope models are represented by open circles and filled circles, respectively. 
The colors of the circles represent the final types of the donors classified in the binary evolution simulations.
The pre-SN states of the RG and AGB models are very close to their final states with remaining-envelope (filled circles) on the H-R diagram.
The pre-SN states of the post-He-flash RG models are displayed by orange squares for comparison with other surviving companion types. Further discussion on the post-He-flash RG surviving companions will take place in Section~\ref{sec:post-He-flash_RG}.
The two dashed lines represent theoretical evolutionary tracks from Figure 6 of \cite{2020ApJ...901...93B} showing two CO WD models with masses of 0.2 $\rm M_{\odot}$ (upper) and 1.3 $\rm M_{\odot}$ (lower), each with a thin H envelope ($q_{\rm H}=10^{-10}$).

In general, core-only models have higher effective temperatures ($10^4 - 2 \times 10^5$ K), a wider range of luminosity values, and generally lower brightness compared to the remaining-envelope models. 
In contrast, the remaining-envelope models maintain a consistent temperature with the pre-SN donor models but a slightly dimmer luminosity level. These characteristics persist for at least 10,000 years after the SN event.
At the end of the simulations at $t_{\rm pe}=100,000$ years, one of the red giant models has evolved to the planetary nebula phase.
For the core-only models with different core types, the RG core models (red open circles) are located within a luminosity range of $\sim$1-100 $\rm L_{\odot}$ and tend to converge to a narrower range between 100 and $<$10,000 years.
The dispersed distribution in the first few ten years likely results from the influence of the MESA mass-loss mechanism in our toy model.
The AGB and TP-AGB core-only models (blue and green open circles) tend to cluster into three distinct groups, maintaining high effective temperature and relatively high luminosity ($\sim$ 20-4,000 $\rm L_{\odot}$) during the 100 to 100,000 year period.

We note that the evolutionary tracks of our core-only models in Figure~\ref{fig:hr_diagram} generally following the direction of the WD cooling tracks from \cite{2020ApJ...901...93B} but still maintain a higher luminosity and temperature distribution. Most of these tracks converge toward the white dwarf branch within $10^5$ years. In addition, their positions lie between the constant-radius contours of $R=0.01 R_\odot$ and $0.1 R_\odot$, suggesting that these surviving cores contract to radii comparable to those of canonical white dwarfs during their cooling evolution. This distribution provides support for identifying these objects as young WDs, although their luminosities remain somewhat elevated relative to old, isolated WDs. The mass distribution of core-only models can be found in the lower panel of Figure~\ref{fig:M_logR}.

\begin{figure*}
\epsscale{1.2}
\plotone{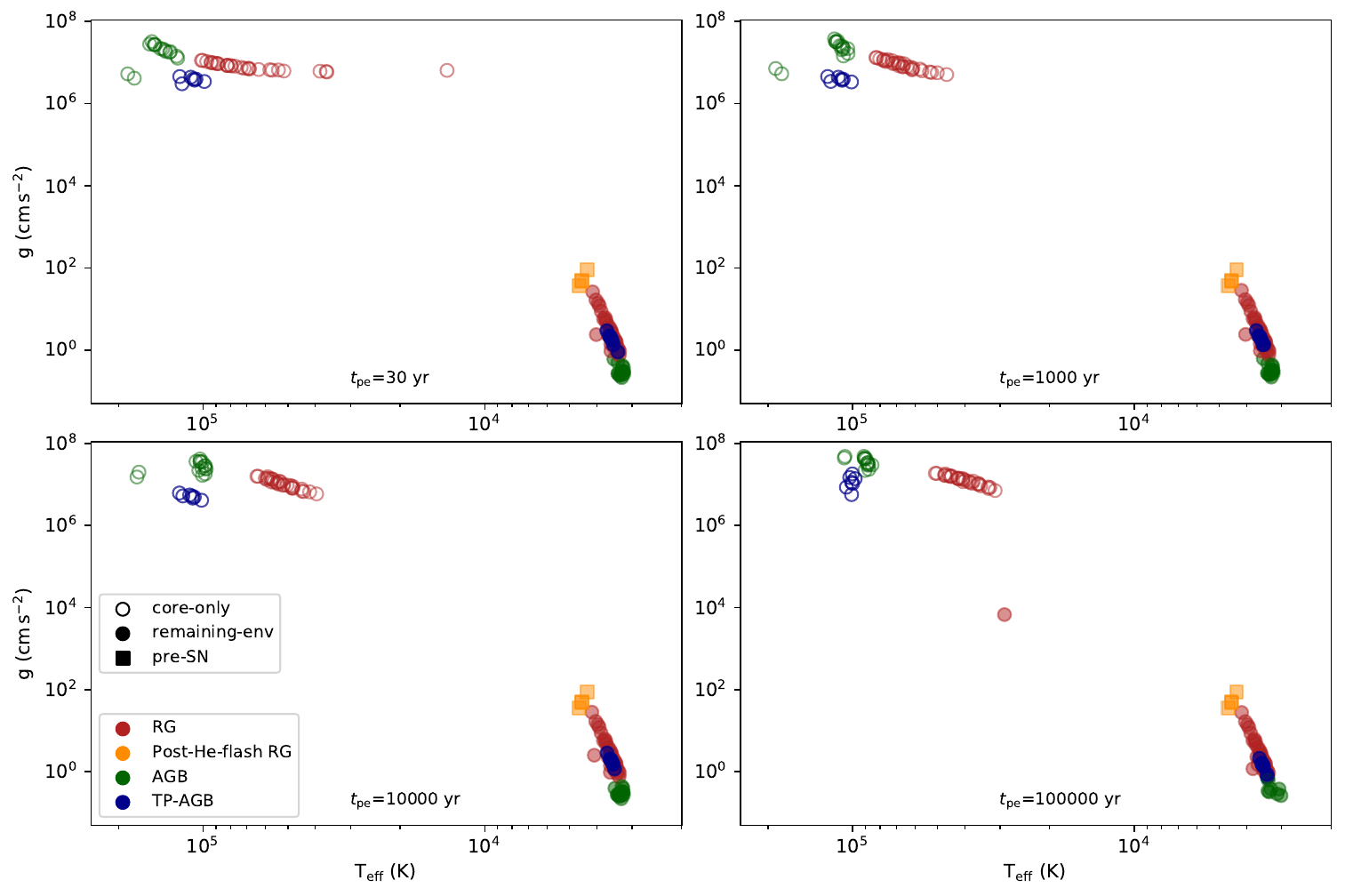}
\caption{Same as Figure~\ref{fig:hr_diagram} but showing surface gravity in the y-axis.}
\label{fig:g_T}
\end{figure*}

Figure~\ref{fig:g_T} has similar settings as in Figure~\ref{fig:hr_diagram} but displays the surface gravity in the y-axis. The surface gravity distribution is more distinctively separated into two groups: the core-only models
are located in the range of $10^{6}-10^8$ $\rm cm \, s^{-2}$, and the remaining-envelope models are distributed within $10^{-1}-10^2$ $\rm cm \, s^{-2}$.

In brief, a higher effective temperature and much higher surface gravity can serve as indicators for differentiating a surviving stellar core companion from nearby stellar population. However, for a companion with a thin layer of envelope, the companion may remain in similar color and brightness to the nearby stellar population, making it challenging to be distinguished in the SNR. 

\section{Discussions}
\label{sec:dis}

\subsection{The Absence of Detectable Surviving Companions in Type Ia SNe: in the View of Giant Companions}

Our results in Section~\ref{sec:obs} suggest that the absence of a companion candidate in the Type~Ia SNRs can also be explained by either not considering hot dwarf-like stars in the searches for the companion, or the surviving companion being indistinguishable from the surrounding stellar population.
For example, \cite{2018ApJ...862..124R} conducted a deep search (detection limit down to $2.6 \, \rm L_{\odot}$) for the surviving companion in Kepler's SNR. They found that the stars within the fields are mostly giants and sub-giants; no MS stars were detected. By exploring parameters, including proper motion and stellar properties, they concluded that no companion was found because no star showed any peculiarity.
However, their method of obtaining stellar properties was limited to a range of effective temperature, $T_{\rm eff}=3,000 - 9,000\,\, \rm K$, which does not cover the higher temperature range of our surviving core-only models. 
They also excluded stars with non-significant ($<$8$\sigma$) proper motion based on the predictions of previous theoretical models, which mainly focused on MS and He star surviving companions. In addition, the selected candidates were found no peculiarity in their stellar properties, but those properties (e.g., $L$, $T_{\rm eff}$, $\log g$) are comparable in scale to those of our remaining-envelope models. Therefore, an evolved non-degenerate companion is not excluded in this case.

Regarding the few existing companion candidates of SNe~Ia, \cite{2023ApJ...947...90R} proposed a companion candidate in a SNR formed 7,500 years ago. The candidate is identified based on its unusual proper motion and trajectory matching the direction toward the SNR center and the timing of SN. The obtained stellar properties show that the companion is a M-type dwarf with $T_{\rm eff} \sim 3,800 \,\rm K$ and $\log g \sim 4.46$. The effective temperature matches the low $T_{\rm eff}$ of our remaining-envelope models, while the surface gravity is located between the values of our core-only models and remaining-envelope models, closer to the former one. The expected companion mass (0.44-0.50 $\rm M_{\odot}$) is at the lower end of our companion mass range (see Figure~\ref{fig:binary_result_final}). Although there is no direct match in the stellar properties with our companion models, an evolved donor companion is still a considerable scenario for the candidate's low $T_{\rm eff}$ and mass.

\subsection{Post-He-flash Red Giant Surviving Companions}
\label{sec:post-He-flash_RG}

The post-He-flash red giant companions in SN Ia progenitor systems are a rare type of surviving companions identified in our systematic study (Section~\ref{sec:result_BE}). 
This category forms when SN explosions occur as the companions have experienced a He flash, leading to a significant decrease in the companion's volume and radius (see Figure~\ref{fig:M_logR}). 
In addition to the radius reduction, these surviving companions also feature larger binary separations compared to models with similar companion masses on the population map (see Figure~\ref{fig:binary_result} for initial conditions and Figure~\ref{fig:binary_result_final} for final conditions).

The extremely large binary-separation-to-companion-radius ratio $A/R_*$ in these systems suggests that the total envelope-stripping assumption during the SN-companion interaction may not apply for the post-He-flash RG surviving companions.
Moreover, the large $A/R_*$ ratio indicates that the SN impact on the surviving companion is minimal, resulting in less pronounced changes in appearance, a shorter timescale for such changes, and a smaller momentum kick from the SN ejecta. These factors make searches for this type of surviving companion more challenging.

Figures~\ref{fig:hr_diagram} and \ref{fig:g_T} display the post-He-flash RG models in their pre-SN stages, denoted by squares (note that these models are not evolving with time; they are shown on all four panels only for comparison). 
These models are distributed at the edges of the population and exhibit relatively low luminosity, high surface temperature, and high surface gravity. 
With an insignificant SN impact, they are likely to remain in the same region on these figures as the surviving companions.
Post-He-flash RG models extend the parameter space of surviving companions, connecting them to the remaining-envelope RG models. As a result, they share a similar conclusion with the remaining-envelope models: they are likely to be difficult to distinguish from the local stellar population, though they have much smaller radii than typical RGs.

\section{SUMMARY AND CONCLUSION}
\label{sec:sum&con}

We presented a series of {\tt MESA} simulations for the symbiotic channel within the single degenerate scenario, from binary evolution to post-impact evolution of the surviving companion.

In binary evolution part, we reimplemented the optically thick wind theory and mass-stripping mechanism proposed by \cite{1996ApJ...470L..97H, 1999ApJ...522..487H} in {\tt MESA}. 
With a total of 1,260 models with different initial binary masses and orbital periods, 427 of them successfully results in SNe~Ia.
Four types of donors among the SN~Ia progenitor systems are identified: red giant (RG) stars with a helium (He) core, post-He-flash RG stars with a He core and a He-burning shell, and AGB and TP-AGB stars with a carbon (C) core. 
An initial donor mass of $M_{\rm d, i} > 2.0 \, \rm M_{\odot}$ is the lower limit for donors to evolve into AGB and TP-AGB stars in SN~Ia progenitor systems. 
Models with $M_{\rm d, i} \lesssim 2.0 \, \rm M_{\odot}$ and having experienced He flashes during RLOF mostly cannot form SNe~Ia progenitors because the companions' radius rapidly reduces and becomes smaller than the Roche-lobe radius before their SNe occur.

We considered two scenarios for surviving companions after SN impact: the core-only models and the remaining-envelope models with 3\% of the original envelope.
Using 92 {\tt MESA} simulations, we explore their observability across the surviving RG, AGB, and TP-AGB companions. 
The surviving companions with thin envelopes retain similar surface temperature and brightness for $\sim$10,000 years before evolving toward the WD branch, while fully-stripped models become much dimmer, hotter, and more compact. Additionally, AGB and TP-AGB (CO-core) surviving companions are generally brighter and hotter than RG (He-core) surviving companions.

Our findings suggest a plausible explanation for the absence of surviving companions in Type Ia SN remnants under the single-degenerate symbiotic channel, which contributes $\lesssim$10\% of SNe~Ia \citep{2023BASBr..34..170L}.
If the donor's envelope is fully stripped by SN impact, the surviving stellar core can be identified by high surface temperature ($10^4 - 2 \times 10^5$ K) and surface gravity ($10^5-10^8$ $\rm cm \, s^{-2}$) without strong peculiar surface composition.
In contrast, if a thin envelope remains, the surviving companion retains pre-SN observational features, making detection challenging.
Our conclusions rely on the optical thick model \citep{1996ApJ...470L..97H, 1999ApJ...522..487H}, and alternative scenarios, such as double-degenerate symbiotic binaries \citep{2010ApJ...719..474D} or the common-envelope-wind model \citep{2017MNRAS.469.4763M}, may yield different outcomes, which we plan to explore in future work.

\begin{acknowledgments}
Y.H.W. acknowledges valuable discussions with Shiau-Jie Rau that took place during her studies at National Tsing Hua University.
This work is supported by the Taiwan National Science and Technology Council (NSTC) through grants NSTC 111-2112-M-007-037, 112-2112-M-007-040 and 113-2112-M-007-031, by the Center for Informatics and Computation in Astronomy (CICA) at National Tsing Hua University through a grant from the Ministry of Education of Taiwan, and by the Center for Theory and Computation at National Tsing Hua University through Grant 112H0001L9. The simulations and data analysis were conducted on the CICA Cluster at National Tsing Hua University and the Taiwania supercomputer at the National Center for High-Performance Computing (NCHC). The analysis and visualization of the simulation data were performed using the analysis toolkit {\tt yt}.

\software{MESA r12115 \citep{2011ApJS..192....3P, 2013ApJS..208....4P, 2015ApJS..220...15P, 2018ApJS..234...34P, 2019ApJS..243...10P}, FLASH 4.5 \citep{2000ApJS..131..273F, 2008PhST..132a4046D}, yt \citep{2011ApJS..192....9T}, \href{https://github.com/jschwab/python-helmholtz}{python-helmholtz} \citep{2000ApJS..126..501T}, Matplotlib \citep{2007CSE.....9...90H}, NumPy \citep{2011CSE....13b..22V}, and SciPy \citep{2019zndo...3533894V}.}
\end{acknowledgments}


\end{document}